\documentclass[aps,pfb,reprint,citeautoscript]{revtex4-1}
\DeclareMathSizes{9}{9}{4}{4}
\DeclareMathSizes{8}{8}{3}{3}

\setcitestyle{super}

\usepackage{filecontents}
\usepackage{hyperref}
\usepackage{graphicx}
\usepackage{comment}
\usepackage{amsmath}
\usepackage{bm}
\usepackage{cases}
\usepackage[euler]{textgreek}
\usepackage[skip=2pt]{caption}
\usepackage[top=0.85in, bottom=0.85in, left=0.7in, right=0.7in]{geometry}
\usepackage{ragged2e}
\usepackage{natbib}
\usepackage{titlesec}
\usepackage{color}

\DeclareCaptionFormat{myformat}{\fontsize{8}{10}\selectfont#1#2#3}
\DeclareCaptionJustification{myjustification}{\justifying}
\hypersetup{colorlinks,linkcolor=black,citecolor=black,urlcolor=black}
\titlespacing*{\section}{0pt}{0.4cm plus0mm minus0mm}{0.4cm plus0mm minus0mm}

\newcommand{\smallpm}{\raisebox{.3\height}{\scalebox{.7}{ $\pm$ }}}

\begin{document}
  \title{Nonlinear thermoelectric response due to energy-dependent transport properties of a quantum dot}
 \author{Artis Svilans}
 \author{Adam M. Burke}
 \author{Sofia Fahlvik Svensson}
 \author{Martin Leijnse}
 \author{Heiner Linke}
  \affiliation{NanoLund and Solid State Physics, Lund University, Box 118, 221 00 Lund, Sweden}

\begin{abstract}\fontsize{8}{11}\selectfont
Quantum dots are useful model systems for studying quantum thermoelectric behavior because of their highly energy-dependent electron transport properties, which are tunable by electrostatic gating. As a result of this strong energy dependence, the thermoelectric response of quantum dots is expected to be nonlinear with respect to an applied thermal bias. However, until now this effect has been challenging to observe because, first, it is  experimentally difficult to apply a sufficiently large thermal bias at the nanoscale and, second, it is difficult to distinguish thermal bias effects from purely temperature-dependent effects due to overall heating of a device. Here we take advantage of a novel thermal biasing technique and demonstrate a nonlinear thermoelectric response in a quantum dot which is defined in a heterostructured semiconductor nanowire. We also show that a theoretical model based on the Master equations fully explains the observed nonlinear thermoelectric response given the energy-dependent transport properties of the quantum dot.
\end{abstract}

\maketitle

\section{Introduction}\fontsize{9}{12}\selectfont

Quantum dots (QDs) are known for their tunable and strongly energy-dependent electron transport properties, which result in a nonlinear response to an applied electrical bias $V_{SD}$. Nonlinear conductance due to the Coulomb blockade \cite{VanHouten1992} is perhaps the most well known example of such nonlinear behavior. It is also well established that the energy-dependent electron transport properties of QDs strongly influence their thermoelectric behavior \cite{Beenakker1992, Staring1993}, which has made them attractive model systems for fundamental studies of quantum thermoelectric effects \cite{Humphrey2002, Edwards1993, ODwyer2006, Esposito2009, Nakpathomkun2010, Jordan2013, Zianni2009}. Nonlinear response to an applied thermal bias $\Delta T$, in particular, has been theoretically investigated in various mesoscopic systems, including resonant tunneling structures \cite{WANG2006, Snchez2013}, multi-terminal quantum conductors \cite{Snchez2013, Meair2013, Whitney2013} and Kondo-correlated devices \cite{Boese2001, Azema2012}. For QDs, one can expect that the quasi-discrete resonance energy spectrum of a QD alone should lead to nonlinear thermoelectric response \cite{Nakpath2010, Svensson2013}. This behavior was explored in detail by Sierra and Sanchez who predicted a strongly nonlinear regime behavior in QDs when $\Delta T$ is about an order of magnitude larger than the background temperature $T_0$ \cite{Sierra2014}.

In experiments, a nonlinear thermovoltage as a function of thermal bias $\Delta T$ has been observed in semiconductor QDs \cite{Staring1993, Svensson2013, Pogosov2006, Hoffmann2009} and in molecular junctions \cite{Reddy2007}. Most recent studies using a tunable thermal bias have shown a strongly nonlinear thermovoltage and thermocurrent in semiconductor nanowire QDs that could not be fully explained by the energy-dependence of the QD resonance energy spectrum alone, and was attributed to a renormalization of resonance energies as a function of heating \cite{Svensson2013}.

The key experimental challenge in the observation of nonlinear thermoelectric behavior in QDs is the ability to apply a tunable and large enough thermal bias $\Delta T$ across a nanoscale object without significant overall heating of the device. The latter can prevent the ability to perform low-temperature experiments, and makes it difficult to distinguish temperature-dependent transport effects from the true nonlinear response to the thermal bias $\Delta T$.

Here, we report measurements of a strongly nonlinear thermocurrent as a function of $\Delta T$ across a QD that is defined by two InP segments within an InAs nanowire. To a large extent the measurements presented here were enabled by a recently developed heater architecture that allows local and electrically non-invasive thermal biasing of a nanowire \cite{Gluschke2014}. This architecture enables tuning of $\Delta T$ over a wide range by applying a relatively small heating power, thus minimizing the parasitic heating effects. We also use theoretical calculations based on Master equations to demonstrate that the experimentally measured thermocurrent can be fully understood from the QD resonance energy spectrum, and is consistent with the previously presented theory in Ref.\hspace{1mm}\cite{Sierra2014}.

\section{Experiment}
\subsection{Device Fabrication}

\begin{figure}[h]
 \includegraphics[width=\columnwidth]{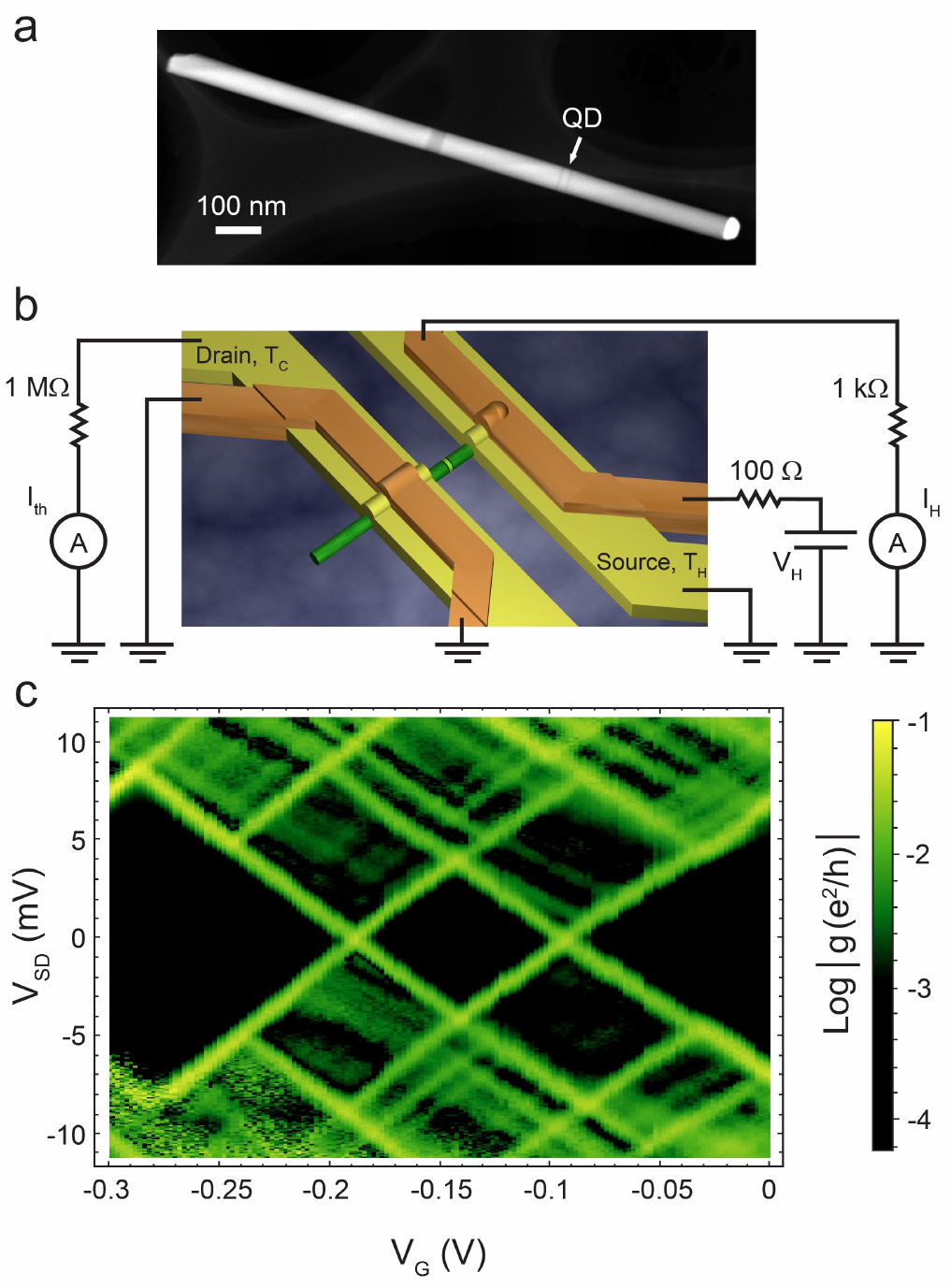}
  \caption{
  (a) Transmission electron microscope image of a nanowire nominally identical to the one used in our thermoelectric device. (b) Device schematic with circuitry diagram for the $I_{th}$ measurement setup. The source and drain contacts in yellow, top-heaters in orange, InAs/InP nanowire in green, quantum dot in light green. The heater over the drain lead is unused. (c) Stability diagram of the InAs quantum dot. Magnitude of differential conductivity, $g=dI/dV_{SD}$, in log$_{10}$-scale as a function of back-gate bias, $V_G$, and source drain bias, $V_{SD}$.
  }\label{fig:1}
\end{figure}

The device consists of a heterostructured InAs/InP nanowire with a 60 nm diameter (see Fig.\hspace{0.5mm}\ref{fig:1}a) that was grown by chemical beam epitaxy seeded by a gold particle \cite{Froberg2008, Persson2007}. Based on transmission electron microscopy (TEM) analyses of $11$ nanowires from the same growth, the InAs/InP nanowire (starting from the seed particle) consists of a $350\smallpm70$ nm InAs segment, followed by a $17\smallpm1.5$ nm long InAs QD defined by two, $4\smallpm3$ nm thick, InP segments, and a second InAs segment of $265\smallpm60$ nm in length. The remaining nanowire, which is not used in the device, consists of a $25$ nm InP plug incorporated for growth reasons and another InAs segment.

The nanowire is contacted to metallic source and drain contacts, as illustrated in Fig.\hspace{0.5mm}\ref{fig:1}b.  Electrically isolated metallic top-heaters pass over the source and drain contacts enabling local dissipation of Joule heat directly on top of the contacts; ensuring heat transfer to the nanowire. Only the heater on top of the source contact was used in the experiments presented here. The device fabrication followed the process developed by Gluschke et al \cite{Gluschke2014}. In brief, electron-beam lithography (EBL) was used to define a pair of source and drain contacts centered around the QD and separated by $300$ nm. A dilute sulfur passivation is performed before source and drain contacts are deposited on the nanowire \cite{Suyatin2007}. A $10$ nm thick layer of HfO$_2$ was deposited via atomic layer deposition to insulate the metallic contacts from the overlying heaters, which were aligned and exposed in a second EBL step. Both the contacts and the heaters were deposited thermally with a metal stack of $25$ nm Ni and $75$ nm Au for the contacts and $25$ nm Ni and 125 nm Au for the heaters. The heater layer was thicker to ensure continuity as the heater steps onto the contact region. The entire device rests on $100$ nm of thermally grown SiO$_2$, allowing the underlying doped Si substrate to be used as a global back gate.

\subsection{Electrical Characterization}

Measurements were conducted in a cryostat in which the estimated electron temperature in the device, $T_0$, was below $1$ K without heating. Bias spectroscopy of the device was carried out using a Stanford Research SRS-830 lock-in amplifier. The voltage from the oscillation output was reduced using a $1:20000$ voltage divider circuit to provide a stable AC source-drain bias amplitude $dV_{SD}=25$ \textmu V $\ll k_B T_0/e$ ($k_B$ - Boltzmann constant, $e$ - elementary charge). To measure the differential conductance $g=dI/dV_{SD}$ as a function of a DC source-drain bias $V_{SD}$, the differential current amplitude, $dI$, was measured in response to $dV_{SD}$, while adding the AC and DC source-drain bias components in a summing box.

\begin{figure*}[ht]
 \includegraphics[width=\textwidth]{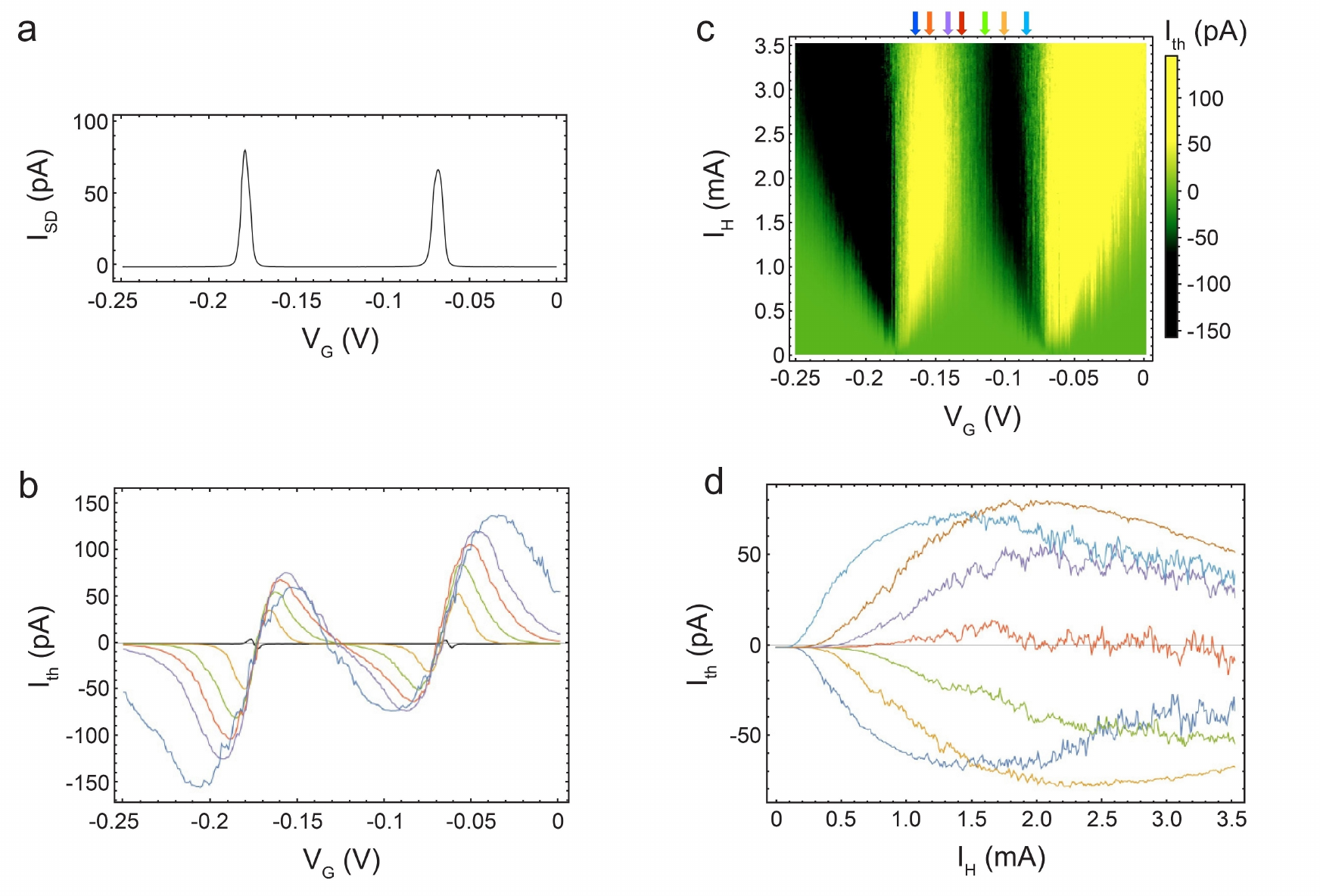}
  \caption{
  (a) Coulomb oscillations in source-drain current $I_{SD}$ as a function of back-gate voltage $V_G$, with the source potential set to $100$ \textmu V. (b) Thermocurrent, $I_{th}$, as a function of back-gate bias for different heater currents $I_H\!=\!(0, 0.35, 0.70, 1.06, 1.41, 3.17 \mbox{ mA})$. (c) Thermocurrent (color) as a function of back-gate voltage, $V_G$, and heating current $I_H$. Arrows along the top correspond to $V_G$ values for traces in (d) as indicated by their color. (d) Thermocurrent as a function of heating current $I_H$ for different $V_G$ values $(-0.165, -0.154, -0.141, -0.131, -0.115, -0.101, -0.085\mbox{ V})$ taken from data in (c).
  }\label{fig:2}
\end{figure*}

To measure Coulomb oscillations (Fig.\hspace{0.5mm}\ref{fig:2}a), a source-drain current, $I_{SD}$, was measured in DC mode using Yokogawa 7651 voltage source to bias the source lead at $100$ \textmu V and a SR570 current preamplifier with $1$ M\textOmega $\mbox{ }$ input impedance.

The set-up used for thermoelectric characterization of the QD nanowire device is shown in Fig.\hspace{0.5mm}\ref{fig:1}b. A thermal bias, $\Delta T$, was applied by running a current $I_H$ through the heater on top of the source contact using a Yokogawa 7651 DC voltage source. The dissipated Joule heat mostly heats the underlying source contact, but is expected to also create a fractional temperature rise in the drain contact \cite{Gluschke2014}. The resulting thermocurrent through the QD nanowire device, $I_{th}$, was amplified via the SR570 current preamplifier.

\subsection{Experimental Results and Discussion}

The QD’s stability diagram, measured as a function of the source-drain voltage, $V_{SD}$, and a back-gate voltage, $V_G$, is shown in Fig.\hspace{0.5mm}\ref{fig:1}c. The dark diamond-like regions represent bias conditions at which the conductivity is suppressed due to Coulomb blockade. From the bias spectroscopy data we estimate a charging energy $E_C$ of $4.0\smallpm0.2$ meV, which is a measure of electron-electron interaction strength in the QD. We also determine the value of the coupling constant $\alpha_G=0.042\smallpm0.04$, which characterizes the capacitive coupling strength between the QD and the back-gate electrode.

Figure \ref{fig:2}b shows $I_{th}$ as a function of $V_G$. The data confirms that our device’s thermoelectric response is typical for QDs \cite{Beenakker1992, Svensson2013, Svensson2012} where $I_{th}$ goes to zero and changes direction at those $V_G$ values where the Coulomb peaks in Fig.\hspace{0.5mm}\ref{fig:2}a are centered. The locations of these thermocurrent zeros do not depend on the heating current, as can be seen in Fig.\hspace{0.5mm}\ref{fig:2}c, which shows $I_{th}$ as a function of $V_G$ and $I_H$. This independence of the $I_{th}$ zeros from $I_H$ is in contrast to previous studies \cite{Svensson2013}, where the nonlinear behavior of $I_{th}$ was strongly influenced by a heating dependent renormalization (shift) of the resonance energies of the QD. The stability of the resonances in the present study is attributed to the benefits of the top-heater architecture where a higher $\Delta T$ can be applied with much less overall background heating of the device \cite{Gluschke2014}.

The core observation of our experiments is the strongly nonlinear behavior of the thermocurrent as a function of $\Delta T$.  This nonlinearity is clearly apparent in Fig.\hspace{0.5mm}\ref{fig:2}d where several back-gate voltage traces, taken from the data in the Fig.\hspace{0.5mm}\ref{fig:2}c, are plotted as a function of $I_H$.

\begin{figure}[h]
 \includegraphics[width=\columnwidth]{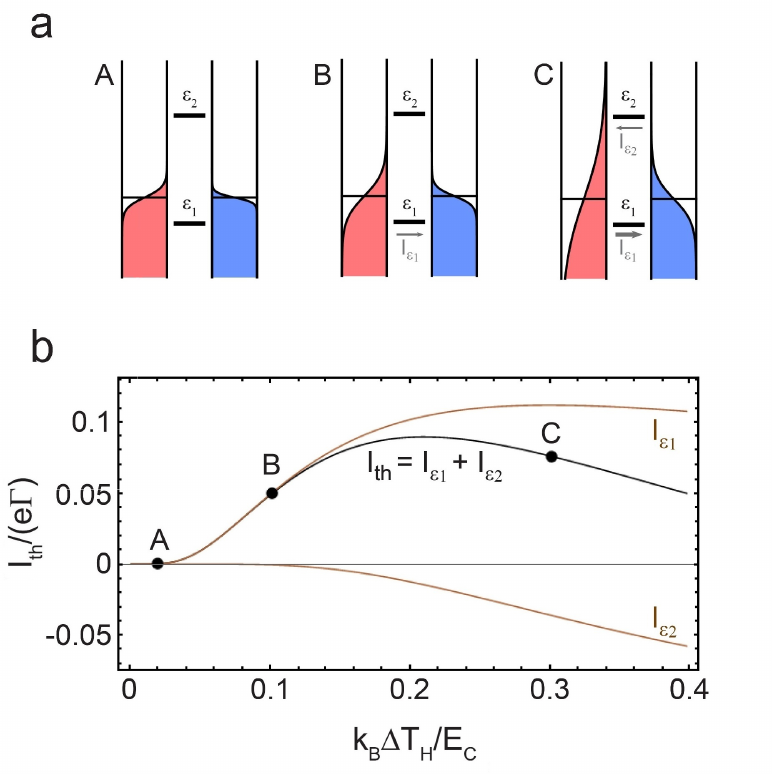}
  \caption{
  (a) Schematic representation of electron distribution in source (red) and drain (blue) leads when the thermal bias is (A) $k_B \Delta T_H/E_C = 0.02$, (B) $0.1$ and (C) $0.3$. Current direction through resonances of a quantum dot is indicated with arrows. Electron energy increases up the vertical axis. (b) Simulated thermocurrent as a function of thermal bias for the back-gate voltage $e\alpha_G V_G/E_C = 0.24$ (black). Brown curves are thermocurrent contributions through each resonance of the quantum dot. See Fig.\hspace{0.5mm}\ref{fig:4} for simulation parameters and Sec.\hspace{0.5mm}\ref{sec:3} for a detailed description.
  }\label{fig:3}
\end{figure}

Several key features can be identified in the observed nonlinear behavior of $I_{th}$, all of which can be understood in terms of the QD’s resonance energy spectrum at different thermal biases. In the following we base our discussion on Ref.\hspace{0.5mm}\cite{Sierra2014} and use phenomenological sketches of a QD resonance spectrum and Fermi-Dirac distributions in the leads to illustrate how the increase in $\Delta T$ can lead to nonlinear effects (Fig.\hspace{0.5mm}\ref{fig:3}). The currents $I_{\varepsilon 1}$ and $I_{\varepsilon 2}$ in Fig.\hspace{0.5mm}\ref{fig:3}b combine to give the overall thermocurrent $I_{th}$ through the QD.

First, we observe that the $I_H$ at which $I_{th}$ starts to rapidly increase depends on $V_G$ (Fig.\hspace{0.5mm}\ref{fig:2}d). As shown in sketch A in Fig.\hspace{0.5mm}\ref{fig:3}a, this behavior can be understood based on the energy of the QD resonances, $\varepsilon_1$ and $\varepsilon_2$. Until the temperature on the hot side reaches a certain value, there is no net current because the electronic states at energies $\varepsilon_1$ and $\varepsilon_2$ in both leads are equally occupied - either completely full or completely empty. This is reflected in point A in Fig.\hspace{0.5mm}\ref{fig:3}b.

The second interesting experimental feature in Fig.\hspace{0.5mm}\ref{fig:2}d is the nonlinear increase of $I_{th}$, as a function of thermal bias. Sketch B in Fig.\hspace{0.5mm}\ref{fig:3}a illustrates how increased heating on the source side leads to a misbalance of the electronic state occupancy in the leads at $\varepsilon_1$. This misbalance leads to a net current as indicated by an arrow in the sketch and by point B in Fig.\hspace{0.5mm}\ref{fig:3}b. Thus, the origin of the nonlinear increase in $I_{th}$ is the nonlinear change of the electronic state occupancy in the leads due to heating.

Finally, $I_{th}$ tends to decrease at higher $I_H$. Ref. \cite{Sierra2014} predicts such behavior due to an increasing backflow of electrons at large thermal bias values $(\Delta T/T\geq 10)$. We believe that the same is true for $I_{th}$ in our experiment, except we expect that we also parasitically heat the drain lead when aiming for high $\Delta T$. Sketch C in Fig.\hspace{0.5mm}\ref{fig:3}a illustrates that the major current contribution, $I_{\varepsilon 1}$, is still provided by the electron transport through $\varepsilon_1$, however, the thermally excited electrons on the source side also leak back through $\varepsilon_2$, thus contributing to the decrease in $I_{th}$. We note that any decrease of the current through $\varepsilon_1$ in the sketch is, in fact, caused by the overall increase in temperature; e.g. slight heating of the drain. However, the backflow of electrons through $\varepsilon_2$ is caused purely by the thermal bias.

\section{Theory}\label{sec:3}
\subsection{Model Description}

We model electron transport through the InAs/InP nanowire by considering a QD which is tunnel-coupled to two electron reservoirs (source and drain leads). Following the experimental setup showed in Fig.\hspace{0.5mm}\ref{fig:1}b the QD is considered in series with a resistive load $R$ to model the input impedance of the current preamplifier. The source and drain leads are characterized by their electrochemical potentials, $u_S = E_F - eV_S$ and $u_D = E_F - eV_D$, where $E_F$ is Fermi energy, and their temperatures, $T_S$ and $T_D$. Electrons in the leads are assumed to occupy states according to the Fermi-Dirac distribution $f_r (E) = \{ 1 + \exp \left[ (E-u_r) / (k_B T_r) \right] \}^{-1}$ and the density of states in the leads is assumed to be a constant. The QD is capacitively coupled to the leads with capacitances $C_S$ and $C_D$, and to the global back-gate with a capacitance $C_G$, giving rise to a charging energy $E_C = e^2 / ( C_S + C_D + C_G )$. In order to model resonance energies we consider a QD in which adding the $N^{th}$ electron changes its state from $i$ to $f$ and that has an electrochemical potential of the form
\begin{equation*}
\mu_{fi}=\epsilon_{fi}+(N-1)E_C- \!\!\!\!\sum\limits_{r=G,S,D} \!\!\!\!\alpha_r V_r.
\end{equation*}
Here $\epsilon_{fi}$ is energy of the single-electron orbital in which the electron is added and $\alpha_r = C_r/( C_S + C_D + C_G )$ are dimensionless coupling constants. We label the probability of the $f^{th}$ state to be occupied $p_f$. Steady-state probabilities for each state occupancy can be represented by a vector $\mathbf{P}$ and are found using the Master equation for a stationary case
\begin{equation*}
  \mathbf{W P}=\mathbf{0}.
\end{equation*}
Here $\mathbf{W}$ is a matrix with elements $W_{fi}$ given by
\begin{equation*}
        W_{fi} = \begin{cases}
                        \sum\limits_{r=S,D}\!\!\left\{\Gamma_{fi}^{r,in}f_r(\mu_{fi})+
                        \Gamma_{fi}^{r,out}\left[1-f_r(\mu_{fi})\right]\right\} \text{,\hspace{2.5mm}if $i\neq f$}\\
                        -\sum\limits_m W_{mf}\text{,\hspace{4.65cm}if $i = f$}
                    \end{cases}
\end{equation*}
where $\mathbf{\Gamma^{S,in}}$, $\mathbf{\Gamma^{D,in}}$, $\mathbf{\Gamma^{S,out}}$ and $\mathbf{\Gamma^{D,out}}$ are matrices containing tunnel rates for single electron tunneling in or out of the QD, involving source or drain leads. Here non-diagonal matrix elements $W_{fi}$ express physical rates at which the QD changes its state from $i$ to $f$. Probability normalization requires that the sum of all occupancy probabilities pf must be $1$.

The current $I_{SD}$ through the QD is then found by adding up current contributions from all possible QD states given the calculated steady state occupancies $p_f$
\begin{equation*}
  I_{SD}=-e\sum\limits_{i,f}p_f \{\Gamma_{fi}^{S,in}f_S(\mu_{fi})-\Gamma_{fi}^{S,out}\left[1-f_S(\mu_{fi})\right]\}.
\end{equation*}
In order to calculate the current $I_{SD}$ through the circuit with the QD and the load $R$ in series, a bias value on the drain side $V_D$ is calculated self-consistently using the Ohms law $V_D=I_{SD}R$.

For the purpose of comparing with our experimental results it is sufficient to consider a QD with only one single electron orbital, in which $N$ can take values 0, 1 or 2. Including electron spin this gives four possible QD states $i,f=\{0, \uparrow, \downarrow, \uparrow\downarrow\}$. In this case, the phenomenological resonance energies $\varepsilon_1$ and $\varepsilon_2$ discussed in the experimental section (Fig.\hspace{0.5mm}\ref{fig:3}) thus correspond to the electrochemical potentials $\mu_{\sigma 0}=\varepsilon_1$ and $\mu_{\uparrow\downarrow\sigma}=\varepsilon_2$, with $\sigma=$ $\uparrow,\downarrow$. For qualitative comparison with experiment we consider the tunnel-barriers to be identical and characterized by a constant tunnel rate $\Gamma$.

\subsection{Simulation Results}\label{sec:3.2}

We now calculate the thermocurrent as a function of temperature in source and drain leads. Since in our experiment the source lead is heated, we label the source temperature $T_S=T_H=T_0+\Delta T_H$ and the drain temperature $T_D=T_C=T_0+\Delta T_C$. In simulations the base temperature $T_0$ is chosen such that $k_B T_0/E_C=0.01$, which is close to the experimental value. Because in the experiments the drain lead is also expected to be somewhat heated we assume $\Delta T_C=\Delta T_H/3$. The ratio between $\Delta T_H$ and $\Delta T_C$ is chosen to obtain a qualitative agreement with the experimental data, but the precise value is not important for the discussed physics.

\begin{figure}[h]
 \includegraphics[width=\columnwidth]{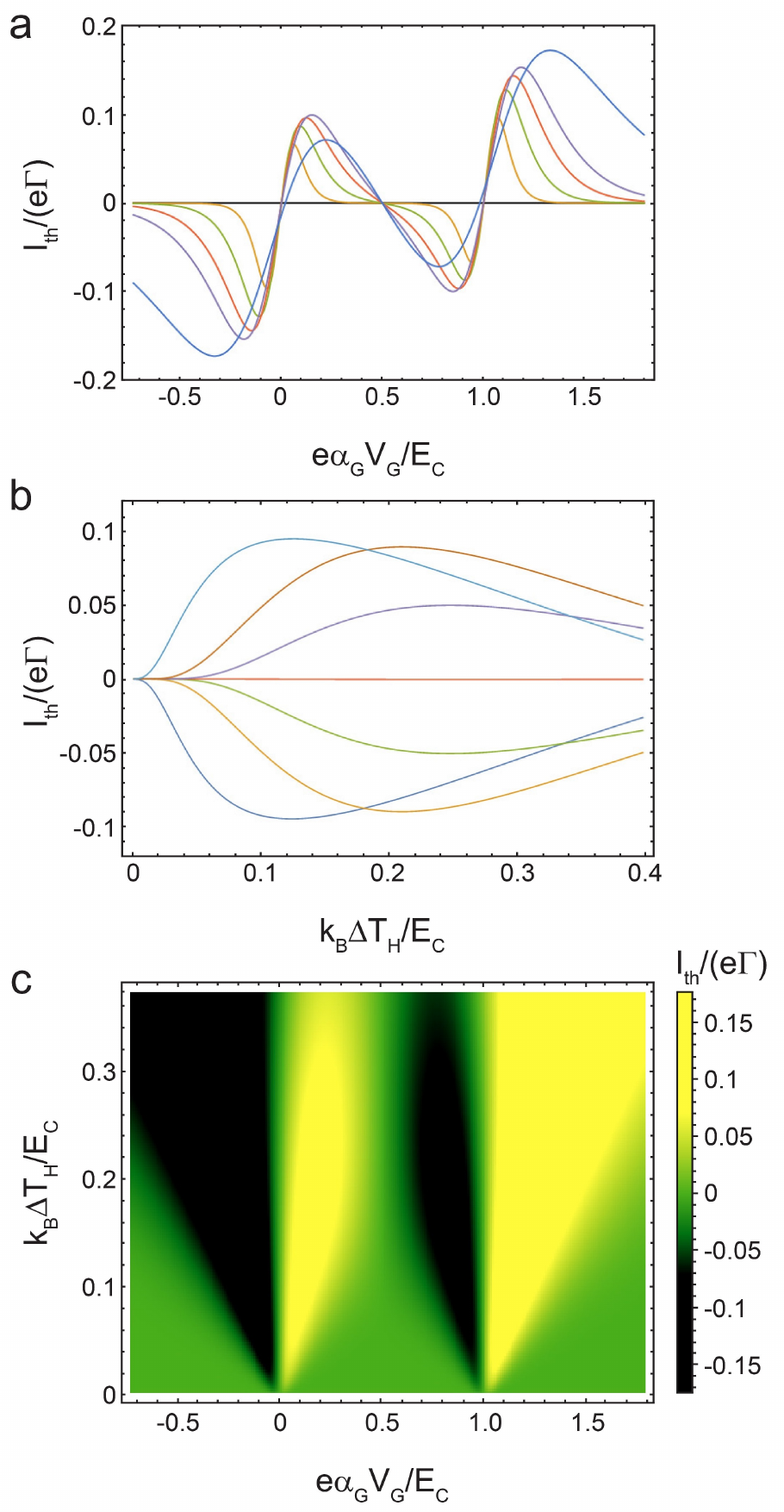}
  \caption{
  (a) Simulated thermocurrent as a function of back-gate voltage for different thermal biases $k_B \Delta T_H/E_C\!\!=\!\!(0, 0.04, 0.08, 0.12, 0.16, 0.32)$. (b) Simulated thermocurrent as a function of the thermal bias for several back-gate voltage values $e\alpha_G V_G/E_C\!\!=\!\!(0.11,0.24,0.37,0.50,0.63,0.76,0.89)$. (c) Simulated thermocurrent (color) as a function of both, back-gate voltage and thermal bias. Other parameters: $\Gamma=5$ GHz, $R=1$ M\textOmega, $T_0=0.01E_C$.
  }\label{fig:4}
\end{figure}

In Fig.\hspace{0.5mm}\ref{fig:4} we sum up our thermocurrent simulation results. Thermocurrent as a function of the back-gate voltage for different thermal bias values is shown in Fig.\hspace{0.5mm}\ref{fig:4}a (compare with the corresponding experimental data in Fig. \ref{fig:2}b). Similarly, we plot the simulated thermocurrent as a function of the thermal bias for different back-gate voltage values in Fig. \ref{fig:4}b. The dimensionless range of thermal bias shown is chosen based on the similarity to Fig.\hspace{0.5mm}\ref{fig:2}d. Finally, the color plot in Fig.\hspace{0.5mm}\ref{fig:4}c is produced using the ranges of the electrochemical potential and the thermal bias used in Figs.\hspace{0.5mm}\ref{fig:4}a and b, and closely matches the experimental result shown in Fig.\hspace{0.5mm}\ref{fig:2}c.

According to our simulations, the source-drain bias $V_{SD}$ that develops across the QD due to the series load at peak currents is estimated to be below $\smallpm 0.04$ $E_C/e$ and therefore does not significantly influence the behavior of the thermocurrent. Note that it is very challenging to measure the temperature in the leads leading up to the QD directly and this was not attempted in the experiment. However, given the qualitative agreement between the experimental thermocurrent data in Fig.\hspace{0.5mm}\ref{fig:2} and the simulated thermocurrent in Fig.\hspace{0.5mm}\ref{fig:4}, one can conclude that the relation between $I_H$ and $\Delta T$ must be close to linear. Moreover, the agreement also suggests that $1$ mA of $I_H$ gives rise to a thermal bias $\Delta T$ of several Kelvin between the source and drain leads.

\section{Conclusions}

In summary, we have reported measurements of a strongly nonlinear thermocurrent in a QD. By comparing our measurements to simulation results, we show that the nonlinear behavior can be fully explained in terms of the QD’s energy-dependent transport properties \cite{Sierra2014}. This is in contrast to earlier experiments \cite{Svensson2013} where this behavior was masked by effects that can also be explained by the overall heating of the device. Our results were enabled by use of a novel heating technique \cite{Gluschke2014} that allows the application of very large $\Delta T$ across a nanoscale device with minimal overall heating of the sample space, even at low temperatures. The ability demonstrated here opens a wide range of quantum thermoelectric experiments in mesoscopic systems.

\acknowledgments
We thank Sebastian Lehmann for the TEM image in Fig.\hspace{0.5mm}\ref{fig:1}a. This work was supported by the People Programme (Marie Curie Actions) of the European Union's Seventh Framework Programme (FP7-People-2013-ITN) under REA Grant agreement no. 608153, by the Swedish Energy Agency (Project P38331-1), by the Swedish Research Council (Project 621-2012-5122) and by NanoLund.
\begin{comment}
\end{comment}
\pagebreak
\bibliography{References}

\begin{thebibliography}{10}
\providecommand{\url}[1]{#1}
\csname url@samestyle\endcsname
\providecommand{\newblock}{\relax}
\providecommand{\bibinfo}[2]{#2}
\providecommand{\BIBentrySTDinterwordspacing}{\spaceskip=0pt\relax}
\providecommand{\BIBentryALTinterwordstretchfactor}{4}
\providecommand{\BIBentryALTinterwordspacing}{\spaceskip=\fontdimen2\font plus
\BIBentryALTinterwordstretchfactor\fontdimen3\font minus
  \fontdimen4\font\relax}
\providecommand{\BIBforeignlanguage}[2]{{%
\expandafter\ifx\csname l@#1\endcsname\relax
\typeout{** WARNING: IEEEtran.bst: No hyphenation pattern has been}%
\typeout{** loaded for the language `#1'. Using the pattern for}%
\typeout{** the default language instead.}%
\else
\language=\csname l@#1\endcsname
\fi
#2}}
\providecommand{\BIBdecl}{\relax}
\BIBdecl

\bibitem{VanHouten1992}
H.~V. Houten, C.~W.~J. Beenakker, and A.~A.~M. Staring, ``Coulomb-blockade
  oscillations in semiconductor nanostructures,'' in \emph{{NATO} {ASI}
  Series}.\hskip 1em plus 0.5em minus 0.4em\relax Springer Science US, 1992,
  pp. 167--216.

\bibitem{Beenakker1992}
C.~W.~J. Beenakker and A.~A.~M. Staring, ``Theory of the thermopower of a
  quantum dot,'' \emph{Phys. Rev. B}, vol.~46, no.~15, pp. 9667--9676, oct
  1992.

\bibitem{Staring1993}
A.~A.~M. Staring, L.~W. Molenkamp, B.~W. Alphenaar, H.~van Houten, O.~J.~A.
  Buyk, M.~A.~A. Mabesoone, C.~W.~J. Beenakker, and C.~T. Foxon,
  ``Coulomb-blockade oscillations in the thermopower of a quantum dot,''
  \emph{Europhysics Letters ({EPL})}, vol.~22, no.~1, pp. 57--62, apr 1993.

\bibitem{Humphrey2002}
T.~E. Humphrey, R.~Newbury, R.~P. Taylor, and H.~Linke, ``Reversible quantum
  brownian heat engines for electrons,'' \emph{Phys. Rev. Lett.}, vol.~89,
  no.~11, aug 2002.

\bibitem{Edwards1993}
H.~L. Edwards, Q.~Niu, and A.~L. de~Lozanne, ``A quantum-dot refrigerator,''
  \emph{Appl. Phys. Lett.}, vol.~63, no.~13, p. 1815, 1993.

\bibitem{ODwyer2006}
M.~F. O'Dwyer, T.~E. Humphrey, and H.~Linke, ``Concept study for a
  high-efficiency nanowire based thermoelectric,'' \emph{Nanotechnology},
  vol.~17, no.~11, pp. S338--S343, may 2006.

\bibitem{Esposito2009}
M.~Esposito, K.~Lindenberg, and C.~V. den Broeck, ``Thermoelectric efficiency
  at maximum power in a quantum dot,'' \emph{{EPL} (Europhysics Letters)},
  vol.~85, no.~6, p. 60010, mar 2009.

\bibitem{Nakpathomkun2010}
N.~Nakpathomkun, H.~Q. Xu, and H.~Linke, ``Thermoelectric efficiency at maximum
  power in low-dimensional systems,'' \emph{Phys. Rev. B}, vol.~82, no.~23, dec
  2010.

\bibitem{Jordan2013}
A.~N. Jordan, B.~Sothmann, R.~S{\'{a}}nchez, and M.~B\"{u}ttiker, ``Powerful
  and efficient energy harvester with resonant-tunneling quantum dots,''
  \emph{Phys. Rev. B}, vol.~87, no.~7, feb 2013.

\bibitem{Zianni2009}
X.~Zianni, ``Thermoelectric efficiency of a quantum dot in the single-electron
  transistor configuration,'' \emph{Journal of Electronic Materials}, vol.~39,
  no.~9, pp. 1996--2001, dec 2009.

\bibitem{WANG2006}
J.~Wang, L.~Wan, Y.~Wei, Y.~Xing, and J.~Wang, ``{Nonlinear} {Thermoelectric}
  {Transport} {Through} a {Double} {Barrier} {Structure},'' \emph{Modern
  Physics Letters B}, vol.~20, no.~05, pp. 215--223, feb 2006.

\bibitem{Snchez2013}
D.~S{\'{a}}nchez and R.~L{\'{o}}pez, ``Scattering theory of nonlinear
  thermoelectric transport,'' \emph{Phys. Rev. Lett.}, vol. 110, no.~2, jan
  2013.

\bibitem{Meair2013}
J.~Meair and P.~Jacquod, ``Scattering theory of nonlinear thermoelectricity in
  quantum coherent conductors,'' \emph{Journal of Physics: Condensed Matter},
  vol.~25, no.~8, p. 082201, jan 2013.

\bibitem{Whitney2013}
R.~S. Whitney, ``Thermodynamic and quantum bounds on nonlinear dc
  thermoelectric transport,'' \emph{Phys. Rev. B}, vol.~87, no.~11, mar 2013.

\bibitem{Boese2001}
D.~Boese and R.~Fazio, ``Thermoelectric effects in kondo-correlated quantum
  dots,'' \emph{Europhysics Letters ({EPL})}, vol.~56, no.~4, pp. 576--582, nov
  2001.

\bibitem{Azema2012}
J.~Azema, A.-M. Dar{\'{e}}, S.~Sch\"{a}fer, and P.~Lombardo, ``Kondo physics
  and orbital degeneracy interact to boost thermoelectrics on the nanoscale,''
  \emph{Phys. Rev. B}, vol.~86, no.~7, aug 2012.

\bibitem{Nakpath2010}
N.~Nakpathomkun, Ph.D. dissertation.

\bibitem{Svensson2013}
S.~F. Svensson, E.~A. Hoffmann, N.~Nakpathomkun, P.~M. Wu, H.~Q. Xu, H.~A.
  Nilsson, D.~S{\'{a}}nchez, V.~Kashcheyevs, and H.~Linke, ``Nonlinear
  thermovoltage and thermocurrent in quantum dots,'' \emph{New Journal of
  Physics}, vol.~15, no.~10, p. 105011, oct 2013.

\bibitem{Sierra2014}
M.~A. Sierra and D.~S{\'{a}}nchez, ``Strongly nonlinear thermovoltage and heat
  dissipation in interacting quantum dots,'' \emph{Phys. Rev. B}, vol.~90,
  no.~11, sep 2014.

\bibitem{Pogosov2006}
A.~G. Pogosov, M.~V. Budantsev, R.~A. Lavrov, A.~E. Plotnikov, A.~K. Bakarov,
  A.~I. Toropov, and J.~C. Portal, ``Coulomb blockade and the thermopower of a
  suspended quantum dot,'' \emph{Jetp Lett.}, vol.~83, no.~3, pp. 122--126, apr
  2006.

\bibitem{Hoffmann2009}
E.~A. Hoffmann, Ph.D. dissertation.

\bibitem{Reddy2007}
P.~Reddy, S.-Y. Jang, R.~A. Segalman, and A.~Majumdar, ``Thermoelectricity in
  molecular junctions,'' \emph{Science}, vol. 315, no. 5818, pp. 1568--1571,
  mar 2007.

\bibitem{Gluschke2014}
J.~G. Gluschke, S.~F. Svensson, C.~Thelander, and H.~Linke, ``Fully tunable,
  non-invasive thermal biasing of gated nanostructures suitable for
  low-temperature studies,'' \emph{Nanotechnology}, vol.~25, no.~38, p. 385704,
  sep 2014.

\bibitem{Froberg2008}
L.~E. Fro?berg, B.~A. Wacaser, J.~B. Wagner, S.~Jeppesen, B.~J. Ohlsson,
  K.~Deppert, and L.~Samuelson, ``Transients in the formation of nanowire
  heterostructures,'' \emph{Nano Letters}, vol.~8, no.~11, pp. 3815--3818, nov
  2008.

\bibitem{Persson2007}
A.~I. Persson, L.~E. Fro?berg, S.~Jeppesen, M.~T. Bjo?rk, and L.~Samuelson,
  ``Surface diffusion effects on growth of nanowires by chemical beam
  epitaxy,'' \emph{J. Appl. Phys.}, vol. 101, no.~3, p. 034313, 2007.

\bibitem{Suyatin2007}
D.~B. Suyatin, C.~Thelander, M.~T. Bj\"{o}rk, I.~Maximov, and L.~Samuelson,
  ``Sulfur passivation for ohmic contact formation to {InAs} nanowires,''
  \emph{Nanotechnology}, vol.~18, no.~10, p. 105307, feb 2007.

\bibitem{Svensson2012}
S.~F. Svensson, A.~I. Persson, E.~A. Hoffmann, N.~Nakpathomkun, H.~A. Nilsson,
  H.~Q. Xu, L.~Samuelson, and H.~Linke, ``Lineshape of the thermopower of
  quantum dots,'' \emph{New Journal of Physics}, vol.~14, no.~3, p. 033041, mar
  2012.

\end{thebibliography}

\end{document}